\documentclass{IEEEtran}
\ifCLASSINFOpdf \else \fi

\usepackage{graphicx}
\usepackage{tikz}
\usepackage[normalem]{ulem}
\usepackage{flushend}

\usepackage[cmex10]{amsmath}
\usepackage{amssymb}

\usepackage[ruled,vlined]{algorithm2e}
\usepackage{capt-of}
\ifCLASSOPTIONcompsoc
 \usepackage[caption=false,font=normalsize,labelfont=sf,textfont=sf]{subfig}
\else
 \usepackage[caption=false,font=footnotesize]{subfig}
\fi
\usepackage{url}
\usepackage{idrislang}
\usepackage{forest}

\usepackage[backend=bibtex]{biblatex}
\begin{document}

\title{Massimult: A Novel Parallel CPU Architecture Based on Combinator Reduction}

\author{
\begin{tabular}[t]{c@{\extracolsep{6em}}c}
J\"urgen Nicklisch-Franken  & Ruslan Feizerakhmanov \\
juergen.nicklisch@symbolian.net & me@russoul.me
\end{tabular}
}


\maketitle


\begin{abstract}
The Massimult project aims to design and implement an innovative CPU architecture based on combinator reduction with a novel combinator base and a new abstract machine. The evaluation of programs within this architecture is inherently highly parallel and localized, allowing for faster computation, reduced energy consumption, improved scalability, enhanced reliability, and increased resistance to attacks.

\par In this paper, we introduce the machine language \textbf{LambdaM}, detail its compilation into \textbf{KVY} assembler code, and describe the abstract machine \textbf{Matrima}. The best part of Matrima is its ability to exploit inherent parallelism and locality in combinator reduction, leading to significantly faster computations with lower energy consumption, scalability across multiple processors, and enhanced security against various types of attacks. Matrima can be simulated as a software virtual machine and is intended for future hardware implementation.
\end{abstract}


\IEEEpeerreviewmaketitle

\section{Introduction}
\IEEEPARstart{V}{on Neumann} architecture relies on the sequential execution of operations. As current CPUs and computers are predominantly built on this architecture, considerable effort has been invested in achieving parallelism within a single core. Modern CPUs are equipped with multiple cores to facilitate parallel processing. However, this approach comes with the trade-off that it requires software specifically designed to leverage these multiple cores effectively.

\par Combinator reduction \cite{Arch}\cite{Graph}\cite{ParComb} represents a fundamentally different approach to computation, providing intrinsic parallel execution without requiring special considerations from the programmer. In this model, a computation can be visualized as a tree, where all branches can evolve concurrently, driven by the leaves, which are composed of combinators and primitive operations. The tree dynamically grows and shrinks as the computation progresses, ultimately reaching completion when no further reductions are possible. However, this tree is effectively a graph, as expressions are shared whenever possible, guaranteed to produce the same result within a functional context.

\par Another key distinction between combinator reduction and the Von Neumann architecture is that, in graph reduction, both program and data are treated uniformly, whereas a Von Neumann architecture separates code and data into different memory regions and handles them differently. In the Von Neumann model, global reads and writes are the typical access patterns, necessitating complex cache hierarchies in modern CPUs to mitigate the inefficiencies associated with global memory access. In contrast, graph reduction operates through the local application of combinators, inherently reducing the need for such extensive caching mechanisms.

\par The Von Neumann architecture and combinator reduction are rooted in distinct theoretical models: the Von Neumann architecture is based on Turing machines, while Combinator Reduction is founded on Lambda Calculus \cite{Barend} and its closely related counterpart, Combinatory Logic \cite{Bimbo}. These underlying models are reflected in different programming paradigms. Imperative programming, which focuses on state changes over time, aligns with the Turing machine model. In contrast, functional programming, centered on the transformation of streams of information, aligns with the principles of Lambda Calculus.

\begin{tikzpicture}[x=0.75pt,y=0.75pt,yscale=-0.8,xscale=0.8]

\draw (333.5,24) -- (332.5,149);

\draw (100,77) node [anchor=north west][inner sep=0.75pt]   [align=left] {Von Neumann Architecture};

\draw (100,30) node [anchor=north west][inner sep=0.75pt]   [align=left] {Turing Machine};

\draw (100,127) node [anchor=north west][inner sep=0.75pt]   [align=left] {Imperative Programming};

\draw (360,77) node [anchor=north west][inner sep=0.75pt]   [align=left] {Massimult Architecture};

\draw (360,30) node [anchor=north west][inner sep=0.75pt]   [align=left] {Lambda Calculus};

\draw (360,127) node [anchor=north west][inner sep=0.75pt]   [align=left] {Functional Programming};

\end{tikzpicture}

\par It is widely recognized that any program can be written in a purely functional style, as exemplified by languages like Haskell or Idris, in a purely imperative style, or, as is often the case, in a hybrid style that combines both paradigms. Traditionally, functional languages are compiled down to imperative assembly languages, reflecting the dominance of Von Neumann architecture in most hardware. However, it is equally feasible to compile imperative languages into a functional assembly language, such as \textbf{KVY} --- section IV-A.

\par Because functional programs avoid side effects and mutable state, they are naturally more amenable to parallel execution, resulting in faster execution times and more efficient use of computational resources compared to their imperative counterparts.

\par Since combinator reduction is based on local operations, it allows for the selective activation of only the parts of a chip that are actively needed at a very fine-grained level. This approach can significantly reduce energy consumption compared to conventional CPUs, which often require larger portions of the chip to be active simultaneously.
Additionally, the intrinsic parallelism of combinator reduction reduces the need to maximize clock frequency, further conserving energy.

\par The functional programming paradigm offers a significant advantage due to its strong theoretical foundations, which facilitate reasoning about program behavior. This advantage extends not only to human developers but also to software tools that can automatically analyze and verify code. Notably, dependently typed functional languages provide a powerful means of encoding software requirements directly within the type system. This enables developers to formally prove that their software adheres to these requirements, paving the way for the development of bug-free, secure, and highly reliable software of unprecedented quality.

\par In this project, we combine deep research with strong engineering skills. We have successfully defined a new combinator code and developed a novel reduction mechanism for parallel and speculative evaluation. Our work includes the implementation of the Matrima virtual machine on a CPU, as well as the development of a \textbf{LambdaM} compiler and interpreter written in \texttt{Idris2}. Our next objective is to achieve competitive performance through GPU emulation and to better understand code optimization in the context of speculative evalation.

\par While the theoretical benefits of our approach are promising, the project will truly gain traction when we demonstrate that this methodology can achieve high effectiveness and speed in practical applications.

\par In this paper, we will first discuss the machine language \textbf{LambdaM} and its compilation into pure lambda calculus. We will then explore the \textbf{KVY} assembler code and, finally, describe the architecture and functioning of the Matrima machine. Readers only interested in novelties can start reading \ref{VComb}

\section{Related Work}

\par The concept of graph reduction as a model for computation has its roots in the 1970s and 1980s, with early research exploring the potential of functional programming languages and graph reduction techniques to enable parallel computation. During this period, a variety of experimental architectures were developed to investigate these ideas, such as the SKIM (S, K, I Machine) \cite{SKIM} and the G-machine \cite{GMachine}, which were specifically designed to execute combinatory logic and lambda calculus efficiently. These early efforts laid the groundwork for the exploration of parallelism in functional programming.

\par In more recent years, projects like the Reduceron \cite{Reduceron2008} \cite{Reduceron2010} and the HVM (Higher-Order Virtual Machine) have expanded on these foundational concepts to explore more efficient implementations of graph reduction for functional programming languages. The Reduceron, for instance, is a hardware accelerator that uses FPGA technology to execute graph reduction more efficiently, optimizing the execution of functional languages like Haskell by mapping combinator graph reductions directly to hardware operations. This approach significantly improves execution speed compared to traditional software implementations.

\par The HVM (Higher-Order Virtual Machine) represents a contemporary effort to optimize the performance of functional languages using an innovative runtime system. Unlike traditional virtual machines, HVM leverages a combination of beta-optimal graph reduction techniques and garbage-collection-free memory management to provide a high-performance environment for executing functional programs. HVM is designed to exploit modern hardware capabilities, aiming to deliver a scalable and energy-efficient runtime that can outperform conventional interpreters and virtual machines for functional languages.

\par Our project builds on these prior efforts by proposing a new combinator code and reduction machinery specifically designed for parallel and speculative evaluation. We aim to advance the state-of-the-art in functional programming and combinator reduction by leveraging GPU emulation and exploring novel hardware implementations. By drawing from both historical and contemporary work in the field, we seek to create a competitive and energy-efficient architecture that enhances the capabilities of functional programming languages.

\section{Machine language and compilation}

\subsection{The Typed Machine Language LambdaM}

\par LambdaM is a functional machine language specifically designed to facilitate the compilation of fully-featured functional languages, such as Haskell, Idris, or Agda. Unlike traditional imperative assembly languages, LambdaM is less technical in nature; instead, it serves as a highly abstract core language that retains the expressive power and functional characteristics of higher-level functional programming paradigms.

\par The primary goal of LambdaM is to provide a minimalistic yet expressive language that preserves the semantics of functional programming while allowing for efficient execution on current and future hardware. LambdaM is typed, ensuring that programs written in the language adhere to strict type rules, which aids in both correctness and optimization during compilation and execution. This typed approach also allows for advanced type-checking mechanisms that can prevent common runtime errors and enforce invariants at the machine level.

\par Technically, LambdaM is based on the Hindley-Milner typed lambda calculus \cite{Hindley}\cite{Milner}, enriched with features such as algebraic data types, \texttt{letrec} expressions, \texttt{case} expressions, \texttt{if-then-else} expressions, and pattern matching. Additionally, LambdaM includes primitive types, literal representations for certain types, primitive functions, and type annotations. The syntax is designed to be similar to functional languages such as Haskell, Idris, or Agda. Readers primarily interested in the novel contributions may choose to skip this section.

\par The use of a typed language in LambdaM is intentional, as it eliminates a wide range of potential run-time errors by enforcing type safety. The decision to adopt a restricted and straightforward Hindley-Milner type system provides the advantage of automatic type inference, which is important for a machine language.

\par A top-level binding in LambdaM defines either a data type or a function. Top-level data type definitions specify the structure and constructors of algebraic data types, providing a foundation for constructing complex data structures. Function bindings define the computational logic, consisting of function names, parameters, type annotations, and bodies composed of expressions and patterns.

\subsubsection{Data Types}

\par A data type definition in LambdaM begins with the \verb!data! keyword, followed by an uppercase identifier that names the newly defined type. This identifier is optionally followed by a sequence of lowercase identifiers, which represent type variables for parametric polymorphism. For example, the definition \verb!List! \verb!a! introduces a generic type \verb!List!, which can be instantiated as \verb!List! \verb!Int! or \verb!List! \verb!String!, depending on the specific type used in place of \verb!a!.

\par The use of type variables allows for the creation of highly versatile and reusable data structures, supporting a wide range of applications while maintaining type safety. This approach enables developers to define generic data types that can be used consistently across different contexts without sacrificing the benefits of a statically-typed system.

\par After the equality sign \verb!=!, the definition proceeds with an uppercase type constructor, followed by a potentially empty sequence of type expressions. These type expressions can be uppercase type names, type variables, or function (arrow) types. This sequence effectively defines a record with types, where the record elements are accessed by their position rather than by name.

\par Furthermore, since the newly defined type name can be referenced within its own definition on the right-hand side, recursive types can be defined. This capability is essential for defining complex data structures such as linked lists, trees, and other recursive constructs that are fundamental in functional programming.

\par It is possible to define multiple constructors for a data type in LambdaM. When doing so, the vertical bar \verb!|! is used as a separator between constructors. This allows instances of the data type to take on different forms, making the type more flexible and expressive. The combination of parametric polymorphism, records, unions (multiple constructors), and recursion forms what is known as an algebraic data type (ADT), a fundamental concept in the functional programming community.

\par Below are three example data type definitions demonstrating these concepts:

\begin{Verbatim}[commandchars=\\\{\}]
\IdrisKeyword{data} \IdrisType{List} \IdrisImplicit{a} = \IdrisData{Nil} | \IdrisData{Cons} \IdrisImplicit{a} (\IdrisType{List} \IdrisImplicit{a}) \\
\IdrisKeyword{data} \IdrisType{IORes} \IdrisImplicit{a} = \IdrisData{MkIORes} \IdrisImplicit{a} \IdrisType{World} \\
\IdrisKeyword{data} \IdrisType{IO} \IdrisImplicit{a} = \IdrisData{MkIO} (\IdrisType{World} -> \IdrisType{IORes} \IdrisImplicit{a})
\end{Verbatim}

\par LambdaM includes built-in types that are typically constructed from literals. Additionally, it provides a set of Prelude types that are immutable and cannot be modified by the user. These Prelude types are fundamental to the language for several reasons: they may have literal representations, they are integral to language constructs, or they are employed by primitive functions, or a combination of these reasons.

\par Furthermore, LambdaM includes standard library types designed to offer efficient implementations of commonly used data structures, such as sequences and maps. These library types are optimized for performance and are essential for developing robust and efficient functional programs. For a detailed overview of these types and their implementations, see Appendix 1.

\subsubsection{Functions}

\par In LambdaM, function definitions begin with a lowercase name, followed by a potentially empty sequence of patterns, an equality sign, and the function's body. Patterns in function definitions serve as the mechanism for destructuring inputs and can take several forms: lowercase variable names for binding values, underscores \texttt{\_} for matching any value without binding it, or type constructors, each followed by an optional sequence of additional patterns. When the sequence of patterns following a constructor is not empty, it must be enclosed in parentheses along with the constructor to ensure proper grouping and clarity.

\par This pattern-matching capability allows for concise and expressive function definitions, facilitating the handling of complex data structures directly within the function signature. The combination of destructuring with algebraic data types enhances both readability and correctness in functional programs by enabling direct manipulation of data forms.

\par When type constructors are used, multiple lines can be utilized to pattern match on the input of a function. It is essential that all possible input cases are covered; otherwise, an error will be raised when loading the program due to incomplete pattern matching. Pattern matching follows a top-down approach where lines are evaluated in order from top to bottom, with the first matching pattern being selected.

\par The following example demonstrates a function that removes the first occurrence of an element from a list. Note that LambdaM does not support overloaded functions or typeclasses; therefore, the equality function (\texttt{eqFunc}) must be explicitly passed to this polymorphic function:

\begin{Verbatim}[commandchars=\\\{\}]
\IdrisFunction{remove} \IdrisBound{eqFunc} \IdrisBound{ele} (\IdrisData{Cons} \IdrisBound{hd} \IdrisBound{tl}) =
  \IdrisKeyword{if} \IdrisBound{eqFunc} \IdrisBound{ele} \IdrisBound{hd}
    \IdrisKeyword{then} \IdrisBound{tl}
    \IdrisKeyword{else} \IdrisData{Cons} \IdrisBound{hd} (\IdrisFunction{remove} \IdrisBound{eqFunc} \IdrisBound{ele} \IdrisBound{tl})
\IdrisFunction{remove} \_ \_ \IdrisData{Nil} = \IdrisData{Nil}
\end{Verbatim}

\par In this example, the \texttt{remove} function takes three arguments: an equality function \texttt{eqFunc}, the element to be removed \texttt{ele}, and a list. The function pattern matches on the list input. If the list is non-empty, it checks if the head element \texttt{hd} matches the element \texttt{ele} using \texttt{eqFunc}. If they are equal, the function returns the tail \texttt{tl}; otherwise, it recursively constructs a new list, retaining \texttt{hd} and removing \texttt{ele} from \texttt{tl}. The second line handles the base case of an empty list, returning \texttt{Nil}.

\par The terms of the language that constitute the right-hand side of function definitions in LambdaM are standard lambda calculus terms, including bound variables, function applications, and abstractions. These terms allow for the definition of functions in a manner consistent with the foundational principles of lambda calculus.

\par LambdaM also supports \texttt{let} expressions, which enable the definition of local functions that are scoped exclusively within the body term of the \texttt{let} and any other functions defined within the same \texttt{let} expression. This scoping mechanism facilitates modular function definitions and helps manage function visibility.

\par Any function in LambdaM can be recursive simply by invoking its own name within its body. Furthermore, functions defined at the top level and within \texttt{let} expressions can be mutually recursive, achieved by referencing each other by name within their respective definitions.

\par The language includes the conditional \texttt{if-then-else} construct, where the condition must be of type \texttt{Bool}, and the types of the expressions for both the true and false branches must be unifiable. This ensures type consistency across different branches of the conditional.

\par LambdaM also provides \texttt{case} expressions for pattern matching. In a \texttt{case} expression, the type of the match term must be unifiable with the types of all specified patterns, and the types of the resulting expressions for each pattern must also be unifiable. The patterns available in \texttt{case} expressions are identical to those used in function definitions. Additionally, all possible patterns must be covered; otherwise, the coverage checker will raise an error, ensuring exhaustive pattern matching and preventing runtime failures due to incomplete matches.

\par The formal language definition of LambdaM is provided below, illustrating the various terms (Tm) that make up the language:

\begin{align*}
Tm &=  \textit{var} && \text{Variable} \\
   &=  Tm_l \, Tm_r && \text{Application} \\
   &=  \lambda \, \textit{var} \, \textbf{.} \, Tm && \text{Abstraction} \\
   &=  \textbf{let} \, \textit{var}_1 \; pat_{1,1} \, \dots \, pat_{1,n} \textbf{=} Tm_1 && \text{Letrec} \\
   &   \quad \quad \quad \dots \\
   &   \quad \quad \; \, \textit{var}_m \; pat_{m,1} \, \dots \, pat_{m,n} \, \textbf{=} Tm_m \\
   &   \quad \; \textbf{in} \, Tm \\
   &=  \textbf{if} \, Tm_c \, \textbf{then} \, Tm_p \, \textbf{else} \, Tm_n && \text{If-Then-Else} \\
   &=  \textbf{case} \, Tm \, \textbf{of} && \text{Case} \\
   &   \quad \quad \quad Pat_1 \rightarrow Tm_1 \\
   &   \quad \quad \quad \dots \\
   &   \quad \quad \quad Pat_n \rightarrow Tm_n \\
   &=  \textit{TypeConstr} && \text{Type Constructor}\\
   &=  \textit{prim} && \text{Primitive}\\
   &=  \textit{literal} && \text{Literal}
\end{align*}

\begingroup
\vspace*{-\baselineskip}
\captionof{figure}{LambdaM Terms}
\vspace*{\baselineskip}
\endgroup

\par Immediately preceding a function definition at the top level or within a \texttt{let} expression, a type annotation can be provided in the form \verb!functionId : typeExpr!. Type annotations are particularly useful in cases involving literal constants, such as \verb!maxSize = 3!, where the type checker may not be able to infer the type on its own.

\par At the beginning of a file, one or more import statements can be included. These statements take the form of \verb!import StdLib!. LambdaM does not support namespaces, so when a simple import statement is used, all top-level names must be distinct to avoid naming conflicts, or you have to qualify the name with the module name, such as \verb!StdLib.Map.Map! or \verb!StdLib.Map.mapInsert!.

\par Finally, there must be a special function defined at the top level called \verb!main!, which serves as the entry point of the program. The \verb!main! function is mandatory for program execution and defines the initial action to be performed when the program starts.

\subsection{Compilation to Pure Untyped Lambda Calculus}

\par In this section, we provide a brief overview of the compilation process utilized by the LambdaM compiler, although the foundational concepts are well established within the functional programming research community. Our novel contributions begin in the subsequent sections.

\par The LambdaM language is compiled into an intermediate representation, referred to as \textbf{LambdaBase}. This intermediate language is based on the pure untyped lambda calculus, augmented with primitive operations and a special construct for recursion. This approach facilitates the efficient translation of LambdaM programs while preserving the core functional properties of the original source code.

\par When loading a source file, the compiler first parses only the import statements. It searches for all required files in the current directory, within the \texttt{Massimult/LambdaM} path, and in any directory specified by the \texttt{LAMBDAM} environment variable. If a required file cannot be found, or if a cyclic dependency is detected, the compiler throws an error. Otherwise, all files are successfully loaded and processed as if they were a single, unified file with all names fully expanded.

\par It is worth noting that future versions of the compiler may implement caching mechanisms to store type-checking and compilation results for source files, thereby avoiding redundant type-checking and compilation for files that have not changed.

\par When a program is loaded, it must first be syntactically correct; otherwise, a lexing error will be raised. Following this, the program must conform to the rules that define valid expressions within the language. If it does not, a parsing error will be generated. Next, all patterns in function definitions and \texttt{case} expressions must be exhaustive; failure to do so will result in a coverage checker error. Finally, the program must pass type checking to ensure that all expressions conform to the expected types; if not, a type error will be thrown.

\subsubsection{Type Checking}

\par Type checking in LambdaM ensures that a program is free from type-related errors, thereby guaranteeing that certain classes of runtime errors cannot occur. While the type checker does not guarantee program termination or prevent out-of-memory errors, it does provide assurances that the program adheres to the specified type constraints. This means that, aside from potential errors arising from primitive functions (such as division by zero) or explicit error handling within the program, the execution is free from type-related failures. Thus, type checking provides a foundational level of safety, ensuring that well-typed programs cannot "go wrong" in terms of type consistency.

\par To achieve type safety, the type checker employs the Hindley-Milner type inference algorithm to automatically infer the types of all expressions in a program. The type system utilized strikes a balance between expressiveness and simplicity, providing types that are as expressive as possible without requiring explicit type annotations in the program code. This characteristic is particularly advantageous for a machine language, as it simplifies programming while maintaining robust type safety.

\par Consequently, a program written in LambdaM does not require explicit type annotations. The type checker infers the most general type for every expression, ensuring that all components fit together correctly, thereby preventing a wide range of potential runtime errors. A program that successfully typechecks is also guaranteed to compile without issues.

\par After loading a program into the LambdaM interpreter, the inferred types of top-level expressions can be displayed using the command \verb!:display t! or its shorthand \verb!:d t!. Additionally, to view the type of a specific expression, the \verb!:type <expr>! command or its shorthand \verb!:t <expr>! can be used.

\subsubsection{Unused Code Removal}

\par Although the type checker requires all loaded code to be consistent, the subsequent compilation step involves a dependency analysis that begins from the \texttt{main} function. This analysis identifies and removes any functions and data constructors that are not referenced, effectively eliminating unused code from the program. This process ensures that only the necessary code is included in the final compiled output, optimizing the program's size and performance.

\subsubsection{Literal Replacement}

\par In this step, literals are replaced with their internal representations. For example, the literal number \texttt{3} could represent different types, such as \texttt{Nat}, \texttt{BinNat}, \texttt{Int}, or \texttt{Float}. The type checker determines the appropriate type based on the context, and the literal is subsequently replaced by its corresponding internal representation. For instance, if the literal is determined to be of type \texttt{Nat}, the number \texttt{3} would be represented internally as \verb!S (S (S Z))!. As previously mentioned, this process may necessitate a type annotation in cases where the type cannot be inferred, such as in a definition like \verb!size = 3!.

\subsubsection{Replacing Patterns and General Case Expressions with Simple Case Expressions}

\par Complex patterns in LambdaM can appear in various contexts, such as top-level function definitions, \texttt{let} expressions, lambda expressions, and \texttt{case} expressions. These patterns are capable of matching data constructors to an arbitrary depth, allowing for expressive and powerful pattern matching capabilities.

\par In contrast, simple \texttt{case} expressions restrict pattern matching to a depth of 1. This means that they can only match the constructors of a single data type and must include all possible constructors of that data type as cases. The compilation process involves transforming complex patterns into equivalent simple \texttt{case} expressions, ensuring that each simple \texttt{case} matches only at the first level of the data type's constructors.

\par Below is an example illustrating a complex pattern and its transformation into a series of simple \texttt{case} expressions:

\begin{Verbatim}[commandchars=\\\{\}]
-- Complex pattern example
\IdrisFunction{exampleFunc} (\IdrisData{Cons}  \IdrisBound{a} (\IdrisData{Cons} \IdrisBound{b} \IdrisBound{bs})) = \IdrisBound{result}
\IdrisFunction{exampleFunc} \_ = \IdrisBound{default}

-- Transformed to simple case expressions
\IdrisFunction{exampleFunc} \IdrisBound{xs} = \IdrisKeyword{case} \IdrisBound{xs} \IdrisKeyword{of}
  \IdrisData{Cons} \IdrisBound{y} \IdrisBound{ys} => \IdrisKeyword{case} \IdrisBound{ys} \IdrisKeyword{of}
    \IdrisData{Cons} \IdrisBound{z} \IdrisBound{zs} => \IdrisBound{result}
    \IdrisData{Nil} => \IdrisBound{default}
  \IdrisData{Nil} => \IdrisBound{default}
\end{Verbatim}

\par While the concept of transforming complex patterns into simple \texttt{case} expressions is straightforward, devising an efficient algorithm for this transformation is considerably more complex. For a detailed discussion of the algorithm and its optimization strategies, we refer the reader to the existing literature \cite{Patterns}. With this transformation, the code is now prepared for the subsequent step, where data types and simple \texttt{case} expressions are further converted into functions.

\subsubsection{Encoding Data Types}

\par Data types in LambdaM are encoded using the Scott encoding \cite{Jansen2013}. In this encoding scheme, each data constructor and simple \texttt{case} expression is transformed into an ordinary function. This approach allows for a uniform treatment of data types and case expressions as first-class functions within the language.

\par Now we show how a simple enumeration (or sum) type can be transformed. For example, consider the following data type definition for a Boolean type and a function that uses a \texttt{case} expression to handle Boolean values:

\par
\begin{Verbatim}
\IdrisKeyword{data} \IdrisType{Bool} = \IdrisData{False} | \IdrisData{True}
\IdrisFunction{cond} \IdrisBound{b} = \IdrisKeyword{case} \IdrisBound{b} \IdrisKeyword{of}
           \IdrisData{False} => \IdrisData{S} \IdrisData{Z}
           \IdrisData{True}  => \IdrisData{Z}
\end{Verbatim}

\par Using the Scott encoding, each data constructor and the \texttt{case} expression are transformed into ordinary functions. The Boolean data type \IdrisType{Bool} and the function \IdrisFunction{cond} would be encoded as follows:

\begin{Verbatim}
\IdrisFunction{False} = \textbackslash \IdrisBound{f1} \IdrisBound{f2} . \IdrisBound{f1}
\IdrisFunction{True}  = \textbackslash \IdrisBound{f1} \IdrisBound{f2} . \IdrisBound{f2}
\IdrisFunction{cond}  = \textbackslash \IdrisBound{b} . \IdrisBound{b} (\IdrisData{S} \IdrisData{Z}) \IdrisData{Z}
\end{Verbatim}

\par In this transformation:
\begin{itemize}
    \item The constructor \IdrisData{False} is encoded as a function that takes two arguments and returns the first argument. This corresponds to the behavior of selecting the first branch in a \texttt{case} expression when the value is \IdrisData{False}.
    \item The constructor \IdrisData{True} is encoded as a function that takes two arguments and returns the second argument. This reflects the behavior of selecting the second branch in a \texttt{case} expression when the value is \IdrisData{True}.
    \item The function \IdrisFunction{cond} is transformed into a function that applies the Boolean argument \IdrisBound{b} to two possible outcomes, \IdrisData{S} \IdrisData{Z} and \IdrisData{Z}, effectively encoding the \texttt{case} expression as a function application. This transformation demonstrates how conditional branching is handled in a functional setting through the use of first-class functions.
\end{itemize}

\par For a simple record (or product) type, the transformation using Scott encoding proceeds as follows:

\begin{Verbatim}
\IdrisKeyword{data} \IdrisType{Tuple} \IdrisBound{a} \IdrisBound{b} = \IdrisData{MkTuple} \IdrisBound{a} \IdrisBound{b}
\IdrisFunction{fst} (\IdrisData{MkTuple} \IdrisBound{a} \IdrisBound{b}) = \IdrisBound{a}
\IdrisFunction{snd} (\IdrisData{MkTuple} \IdrisBound{a} \IdrisBound{b}) = \IdrisBound{b}
=>
\IdrisFunction{MkTuple} = \textbackslash \IdrisBound{a} \IdrisBound{b} \IdrisBound{f} . \IdrisBound{f} \IdrisBound{a} \IdrisBound{b}
\IdrisFunction{fst} = \textbackslash \IdrisBound{f} . \IdrisBound{f} (\textbackslash \IdrisBound{a} \IdrisBound{b} . \IdrisBound{a})
\IdrisFunction{snd} = \textbackslash \IdrisBound{f} . \IdrisBound{f} (\textbackslash \IdrisBound{a} \IdrisBound{b} . \IdrisBound{b})
\end{Verbatim}

\par In this transformation:
\begin{itemize}
    \item The constructor \IdrisData{MkTuple} is encoded as a function that takes two arguments, \IdrisBound{a} and \IdrisBound{b}, and a function \IdrisBound{f}, and applies \IdrisBound{f} to \IdrisBound{a} and \IdrisBound{b}. This encoding effectively turns the tuple constructor into a higher-order function.
    \item The function \IdrisFunction{fst}, which retrieves the first element of the tuple, is transformed into a function that takes a function \IdrisBound{f} and applies \IdrisBound{f} to a lambda expression that returns the first element \IdrisBound{a}.
    \item Similarly, the function \IdrisFunction{snd}, which retrieves the second element of the tuple, is encoded as a function that takes a function \IdrisBound{f} and applies \IdrisBound{f} to a lambda expression that returns the second element \IdrisBound{b}.
\end{itemize}

\par This transformation demonstrates how product types, like tuples, can be represented in a purely functional setting using functions. By encoding tuples as functions, LambdaM maintains uniformity in its treatment of data types, supporting a higher level of abstraction and functional manipulation.

\par The following example defines a \texttt{List} type, which exhibits both an enumeration (sum) type and a record (product) type. The \texttt{List} type is an enumeration type because it can be either \IdrisData{Nil} (representing an empty list) or \IdrisData{Cons} (representing a non-empty list). It is also a record type because the \IdrisData{Cons} constructor includes both an element and a list of elements as its members:

\begin{Verbatim}
\IdrisKeyword{data} \IdrisType{List} \IdrisBound{a} = (\IdrisData{Cons} \IdrisBound{a} (\IdrisType{List} \IdrisBound{a})) | \IdrisData{Nil}
\IdrisFunction{head} (\IdrisData{Cons} \IdrisBound{a} \IdrisBound{rest}) = \IdrisData{Just} \IdrisBound{a}
\IdrisFunction{head} \IdrisData{Nil} = \IdrisData{Nothing}
\IdrisFunction{tail} (\IdrisData{Cons} \IdrisBound{a} \IdrisBound{rest}) = \IdrisData{Just} \IdrisBound{rest}
\IdrisFunction{tail} \IdrisData{Nil} = \IdrisData{Nothing}
=>
\IdrisFunction{Cons}  = \textbackslash \IdrisBound{a} \IdrisBound{r} \IdrisBound{f1} \IdrisBound{f2} . \IdrisBound{f1} \IdrisBound{a} \IdrisBound{r}
\IdrisFunction{Nil}   = \textbackslash \IdrisBound{f1} \IdrisBound{f2} . \IdrisBound{f2}
\IdrisFunction{head}  = \textbackslash \IdrisBound{l} . \IdrisBound{l} (\textbackslash \IdrisBound{a} \IdrisBound{r} . \IdrisData{Just} \IdrisBound{a}) \IdrisData{Nothing}
\IdrisFunction{tail}  = \textbackslash \IdrisBound{l} . \IdrisBound{l} (\textbackslash \IdrisBound{a} \IdrisBound{r} . \IdrisData{Just} \IdrisBound{r}) \IdrisData{Nothing}
\end{Verbatim}

\par In this transformation:
\begin{itemize}
    \item The constructor \IdrisData{Cons} is encoded as a function that takes an element \IdrisBound{a}, a list \IdrisBound{r} (the rest of the list), and two additional function arguments, \IdrisBound{f1} and \IdrisBound{f2}. It applies \IdrisBound{f1} to \IdrisBound{a} and \IdrisBound{r}, effectively representing the non-empty list case.
    \item The constructor \IdrisData{Nil} is represented as a function that takes two function arguments, \IdrisBound{f1} and \IdrisBound{f2}, and returns \IdrisBound{f2}, corresponding to the empty list case.
    \item The function \IdrisFunction{head} retrieves the first element of the list. It is transformed into a function that applies the list \IdrisBound{l} to two lambdas: one that returns \IdrisData{Just} \IdrisBound{a} when the list is non-empty, and \IdrisData{Nothing} when it is empty.
    \item Similarly, the function \IdrisFunction{tail} retrieves the rest of the list (excluding the first element). It is encoded as a function that applies the list \IdrisBound{l} to two lambdas: one that returns \IdrisData{Just} \IdrisBound{r} for a non-empty list and \IdrisData{Nothing} for an empty list.
\end{itemize}

\par This transformation illustrates how a combined enumeration and record type, such as a \texttt{List}, can be encoded in a purely functional style using Scott encoding. By converting data constructors and functions into first-class functions, LambdaM maintains a consistent and flexible representation for complex data structures.

\par This encoding closely aligns with the conventional handling of data types while maintaining efficiency. Scott encoding provides a functional representation of data constructors and case expressions, transforming them into first-class functions. The general transformation for a data type and its corresponding \texttt{case} expression is formalized as follows:

\begin{align*}
&\text{data} \, N \, = \, C_1 \, v_{1,1} \dots v_{1,n} \, | \, \dots \, | \, C_m \, v_{m,1} \dots v_{m,n} \\
&\quad \quad \quad \Downarrow \\
&C_1 = \lambda \, v_{1,1} \dots v_{1,n} \, f_1 \dots f_m \, . \, f_1 \, v_{1,1} \dots v_{1,n} \\
&\quad \quad \quad \dots \\
&C_m = \lambda \, v_{m,1} \dots v_{m,n} \, f_1 \dots f_m \, . \, f_m \, v_{m,1} \dots v_{m,n} \\
&\\
&\text{case} \, x \, \text{of} \\
&\quad C_1 \, v_{1,1} \dots v_{1,n} \Rightarrow b_1 \\
&\quad \quad \quad \dots \\
&\quad C_m \, v_{m,1} \dots v_{m,n} \Rightarrow b_m \\
&\quad \quad \quad \Downarrow \\
&\lambda \, f \, . \, f \, (\lambda \, v_{1,1} \dots v_{1,n} \, . \, b_1) \\
&\quad \quad \quad \dots \\
&\quad \quad \quad (\lambda \, v_{m,1} \dots v_{m,n} \, . \, b_m) \\
\end{align*}

\begingroup
\vspace*{-\baselineskip}
\captionof{figure}{Scott Encoding}
\vspace*{\baselineskip}
\endgroup

\par In this general transformation:
\begin{itemize}
    \item Each data constructor \(C_i\) is transformed into a function that takes its arguments \(v_{i,1}, \dots, v_{i,n}\) and a series of functions \(f_1, \dots, f_m\). The constructor function applies the corresponding function \(f_i\) to its arguments, encoding the selection logic of case expressions.
    \item The \texttt{case} expression is transformed into a lambda function that takes a function \(f\) and applies it to a series of lambda expressions. Each lambda expression represents a branch of the original \texttt{case} expression, effectively encapsulating the branching logic in a functional form.
\end{itemize}

\par This transformation leverages the power of higher-order functions to represent both data structures and control flow, demonstrating the expressive capabilities of the Scott encoding in a functional programming context.

\subsubsection{Replacing \texttt{let} Expressions with Lambdas}

\par Replacing \texttt{let} expressions with lambda expressions is straightforward. A \texttt{let} expression of the form \verb!let x = e in t! is translated into the equivalent lambda expression \verb!(\x. t) e!. This transformation leverages lambda abstraction to achieve the same effect as a local binding.

\par Consequently, in \texttt{let} expressions containing multiple definitions, only those definitions declared earlier can be referenced in subsequent expressions. This restriction also applies to top-level function definitions, which can be viewed as a \texttt{let} expression without an explicitly written \texttt{let}. To enforce this order, the compiler performs a dependency analysis and arranges the defining terms accordingly.

\par However, when mutually recursive functions are present, additional considerations are required. The compilation of such functions is addressed in the next subsection.

\subsubsection{Recursion}

\par In LambdaM, recursion is supported both for single functions and for multiple functions that reference each other. A function is considered recursive if it refers to its own name within its definition. Similarly, functions are considered mutually recursive if two or more functions refer to each other. Both forms of recursion are fully supported in LambdaM, allowing for the definition of complex recursive algorithms.

\par We begin by addressing simple recursion. There are multiple strategies for compiling recursive functions, such as introducing cycles in the evaluation graph or passing the function itself as an extra argument to enable self-reference. In our approach, we utilize a fixpoint combinator. Specifically, we employ an applicative order fixpoint combinator, which we denote as \verb!Y!. The \verb!Y! combinator is defined as \verb!Y f x = f (Y f) x!, which allows the function \texttt{f} to recursively call itself in a purely functional manner.

\par The factorial function is used as an illustrative example to demonstrate the compilation process through multiple transformations. The first transformation involves encoding data types, patterns, and \texttt{case} expressions, as previously described. The second transformation demonstrates how recursion is handled using the \texttt{Y} combinator.

\begin{Verbatim}
-- Original recursive factorial function
\IdrisFunction{fac} \IdrisData{Z} = \IdrisData{S} \IdrisData{Z}
\IdrisFunction{fac} (\IdrisData{S} \IdrisBound{n}) = \IdrisFunction{mul} (\IdrisFunction{fac} \IdrisBound{n}) (\IdrisData{S} \IdrisBound{n})

-- Step 1: Scott Encoding
\IdrisFunction{fac} = \textbackslash \IdrisBound{sn} .
    \IdrisBound{sn} (\textbackslash \IdrisBound{n} . \IdrisFunction{mul} (\IdrisFunction{fac} \IdrisBound{n}) (\IdrisData{S} \IdrisBound{n})) (\IdrisData{S} \IdrisData{Z})

-- Step 2: Using the fixpoint combinator
\IdrisFunction{fac} = \IdrisKeyword{Y} (\textbackslash \IdrisBound{fac'} \IdrisBound{sn} .
    \IdrisBound{sn} (\textbackslash \IdrisBound{n} . \IdrisFunction{mul} (\IdrisBound{fac'} \IdrisBound{n}) (\IdrisData{S} \IdrisBound{n})) (\IdrisData{S} \IdrisData{Z}))
\end{Verbatim}

\par In these transformations:
\begin{itemize}
    \item \textbf{Original Function Definition:} The factorial function, \IdrisFunction{fac}, is defined recursively with two cases: one for the base case (\IdrisData{Z}) and one for the recursive case (\IdrisData{S} \IdrisBound{n}).
    \item \textbf{Step 1: Encoding:} The original recursive function is transformed to a lambda expression that encodes pattern matching using function application. The function \IdrisFunction{fac} takes an argument \IdrisBound{sn} representing the structure of the natural number (either \IdrisData{Z} or \IdrisData{S} \IdrisBound{n}) and applies appropriate logic for each case.
    \item \textbf{Step 2: Applying the Fixpoint Combinator:} To handle recursion, we introduce the \texttt{Y} combinator. This transformation replaces the direct recursive call to \IdrisFunction{fac} with a reference to the function itself (\IdrisBound{fac'}) using the fixpoint combinator. This allows recursion to be expressed without explicit self-reference, enabling the factorial calculation to proceed in a purely functional context.
\end{itemize}

\par These steps illustrate the transformation of a recursive function into a form that is compatible with the functional paradigm of LambdaM. By utilizing the \texttt{Y} combinator, we ensure that recursion is managed efficiently within the functional framework, preserving both the expressiveness and correctness of the original program.

\par For mutually recursive functions, we extend the use of the \texttt{Y} combinator by passing a tuple of functions rather than a single function. This approach enables the handling of multiple functions that recursively reference each other, allowing them to be defined together in a cohesive manner. To illustrate this, we consider the classic example of mutually recursive functions, \texttt{even} and \texttt{odd}. Although this example is straightforward and primarily pedagogical, it effectively demonstrates the transformation process involved in managing mutual recursion using the \texttt{Y} combinator:

\begin{Verbatim}
-- Original mutually recursive functions
\IdrisFunction{even} \IdrisData{Z} = \IdrisData{True}
\IdrisFunction{even} (\IdrisData{S} \IdrisBound{n}) = \IdrisFunction{odd} \IdrisBound{n}
\IdrisFunction{odd} \IdrisData{Z} = \IdrisData{False}
\IdrisFunction{odd} (\IdrisData{S} \IdrisBound{n}) = \IdrisFunction{even} \IdrisBound{n}

-- Step 1: Scott Encoding
\IdrisFunction{even} = \textbackslash \IdrisBound{sn} .
  \IdrisBound{sn} (\textbackslash \IdrisBound{n} . \IdrisFunction{odd} \IdrisBound{n}) \IdrisData{True}
\IdrisFunction{odd} = \textbackslash \IdrisBound{sn} .
  \IdrisBound{sn} (\textbackslash \IdrisBound{n} . \IdrisFunction{even} \IdrisBound{n}) \IdrisData{False}

-- Step 2: Handling mutual Recursion
\IdrisFunction{evenodd} = \IdrisKeyword{Y} (\textbackslash \IdrisBound{eo} .
  (\IdrisFunction{Tuple} (\textbackslash \IdrisBound{sn} .
    \IdrisBound{sn} (\textbackslash \IdrisBound{n} . (\IdrisFunction{snd} \IdrisBound{eo}) \IdrisBound{n}) \IdrisData{True})
   (\textbackslash \IdrisBound{sn} .
    \IdrisBound{sn} (\textbackslash \IdrisBound{n} . (\IdrisFunction{fst} \IdrisBound{eo}) \IdrisBound{n}) \IdrisData{False})))

\IdrisFunction{even} = \IdrisFunction{fst} \IdrisFunction{evenodd}
\IdrisFunction{odd}  = \IdrisFunction{snd} \IdrisFunction{evenodd}
\end{Verbatim}

\par In this transformation:
\begin{itemize}
    \item \textbf{Original Function Definitions:} The functions \IdrisFunction{even} and \IdrisFunction{odd} are defined with two cases each: one for the base case (\IdrisData{Z}) and another for the recursive case (\IdrisData{S} \IdrisBound{n}).
    \item \textbf{Step 1: Encoding:} The original recursive functions are transformed into lambda expressions that handle pattern matching through function application. Here, \IdrisFunction{even} and \IdrisFunction{odd} are encoded to apply the appropriate logic based on the structure of the input.
    \item \textbf{Step 2: Using the Fixpoint Combinator:} To handle mutual recursion, we construct a tuple of functions (\texttt{Tuple e o}) representing \IdrisFunction{even} and \IdrisFunction{odd}. The \texttt{Y} combinator is applied to this tuple, allowing both functions to recursively reference each other through the tuple. The functions \IdrisFunction{fst} and \IdrisFunction{snd} are then used to extract the individual functions from the tuple.
\end{itemize}


\subsubsection{LambdaBase Language}

\par With these transformations, the compilation of LambdaM to a representation based on pure untyped lambda calculus, augmented with primitives and recursion, is complete. The resulting intermediate language, which we call \textbf{LambdaBase}, consists of a minimal set of constructs that capture the essence of functional computation while supporting recursion and primitive operations. Below is the compact definition of the \textbf{LambdaBase} language:

\begin{align*}
Tm &=  \textit{var} && \text{Bound Variable} \\
   &=  Tm_l \, Tm_r && \text{Application} \\
   &=  \lambda \, \textit{var} \, . \, Tm && \text{Abstraction} \\
   &=  \underline{Y} && \text{Recursion via Fixpoint Combinator} \\
   &=  \textit{prim} \dots && \text{Primitives} \\
\end{align*}

\begingroup
\vspace*{-\baselineskip}
\captionof{figure}{LambdaBase Language}
\vspace*{\baselineskip}
\endgroup

\par In this definition:
\begin{itemize}
    \item \textbf{Bound Variable (\textit{var}):} Represents a variable that is bound within a lambda abstraction.
    \item \textbf{Application ($Tm_l \, Tm_r$):} Represents the application of one term to another, capturing function application in the calculus.
    \item \textbf{Abstraction ($\lambda \, \textit{var} \, . \, Tm$):} Represents the definition of an anonymous function with a bound variable.
    \item \textbf{Recursion via Fixpoint Combinator ($\underline{Y}$):} Introduces recursion into the language by using a fixpoint combinator, allowing for the definition of recursive functions.
    \item \textbf{Primitives (\textit{prim} \dots):} Represents a set of primitive operations, such as arithmetic functions, that are treated as basic operations in the language.
\end{itemize}

\par The \textbf{LambdaBase} language provides a foundational core for further compilation to combinatory machine code, while preserving the essential features of functional computation.

\subsubsection{And Back Again}

\par When using LambdaM as an interpreter, the compilation process is identical to that of using LambdaM as a compiler. However, in an interpretive context, the output is expected to be presented in the form of LambdaM expressions. Reconstructing LambdaM terms from \textbf{LambdaBase} terms is feasible, but only when the type of the expression is known.

\par The reason for this constraint is that different constructors in LambdaM can be compiled into indistinguishable functions in \textbf{LambdaBase}. Consequently, without type information, it would be impossible to accurately reconstruct the original constructors. Fortunately, for every expression interpreted in LambdaM, the result type is always known, enabling the correct reconstruction of both constructors and literals. Functions are also displayed correctly, although their presentation may not always be optimal at this stage.

\par This ensures that, despite the compilation to a lower-level representation, the interpreter can effectively present results in the original, more readable LambdaM format, preserving both the semantic clarity and usability of the interpreted output.

\section{KVY Combinator Code}

\par The simplest combinator base requires only two combinators, \verb!K! and \verb!S!, as demonstrated by Smullyan \cite{Smullyan}. The reduction rules for these combinators are defined as follows:

\begin{align*}
K \, x \, y \, &\Rightarrow \, x \\
S \, x \, y \, z \, &\Rightarrow \, x \, z \, (y \, z)
\end{align*}

\begingroup
\vspace*{-\baselineskip}
\captionof{figure}{KS reduction rules}
\vspace*{\baselineskip}
\endgroup

\par A computer that implements only the \verb!K! and \verb!S! combinators with the aforementioned reduction rules is computationally universal, or Turing complete. This means it possesses the capability to perform any computation that can be executed by more complex computing machines. This result is remarkable, highlighting the power of a system that is even simpler than the Lambda calculus.

\par Despite their simplicity, the \verb!K! and \verb!S! combinators provide a sufficient foundation for constructing any computable function. This minimalistic approach not only facilitates a deeper understanding of the fundamental principles of computation but also allows for optimizations that can exploit the inherent parallelism in combinator-based systems. Such properties make them highly suitable for theoretical investigations into the nature of computation and for practical applications where simplicity and elegance are desired.

\par The equivalence in expressiveness between combinators and the Lambda calculus can be demonstrated by defining a transformation from one formalism to the other. The transformation from Lambda calculus to combinators is known as \textit{abstraction elimination}, with its core mechanism referred to as \textit{bracket abstraction}. This transformation can be defined as follows, where the transformation is indicated by square brackets:

\begin{align*}
\langle Tm \rangle _x \, &\mapsto \, K \, Tm \quad \text{if} \, x \, \notin \, Tm \\
\langle x \rangle _x \, &\mapsto \, S \, K \, K \\
\langle Tm_l \, Tm_r \rangle _x \, &\mapsto \, S \,  \langle Tm_l \rangle _x \, \langle Tm_r \rangle _x \\
\end{align*}

\begin{align*}
[\lambda \, x \, . \, Tm] \, &\mapsto \, \langle [Tm] \rangle _x \\
[x] \, &\mapsto \, x \\
[Tm_l \, Tm_r] \, &\mapsto \, [Tm_l] \, [Tm_r] \\
[Y] \, &\mapsto \, Y
\end{align*}

\begingroup
\vspace*{-\baselineskip}
\captionof{figure}{KS Abstraction Elimination}
\vspace*{\baselineskip}
\endgroup

\par The reverse transformation is straightforward and involves the application of the reduction rules defined for each combinator.

\par A notable issue with using only the \verb!K! and \verb!S! combinators is the significant increase in the number of combinators compared to the number of lambda expressions, leading to a proliferation of fine-grained operations, which result in inefficient computation performance. Enhancing the combinator base by including additional combinators greatly improves efficiency. A more effective base includes the combinators \texttt{I}, \texttt{K}, \texttt{S}, \texttt{B}, and \texttt{C}, each with its respective reduction rules:

\begin{align*}
I \, x \, &\Rightarrow \, x \\
K \, x \, y \, &\Rightarrow \, x \\
S \, x \, y \, z \, &\Rightarrow \, x \, z \, (y \, z) \\
B \, x \, y \, z \, &\Rightarrow \, x \, (y \, z) \\
C \, x \, y \, z \, &\Rightarrow \, x \, z \, y
\end{align*}

\begingroup
\vspace*{-\baselineskip}
\captionof{figure}{IKSBC reduction rules}
\vspace*{\baselineskip}
\endgroup

\par Early implementations of functional languages that leveraged combinatory logic relied on a fixed set of pre-defined combinators. To enhance code compactness and minimize the number of fine-grained reductions, these implementations introduced additional combinators, e.g. Turner. These combinators served to minimize the growth of the number of combinators with respect to the number of lambda expressions. Our approach advances as high as mapping one lambda to exactly one combinator.

\subsection{V: A Family of Combinators Indexed by a Multipath}\label{VComb}

\par The concept is to introduce a family of combinators, each indexed by a multi-path within a binary tree, which can be interpreted as a binary tree lacking any content. These combinators operate by inserting an element into the specific positions designated by the multi-path within the tree structure. Due to the nature of the multi-path, the element can be inserted one or more times at various locations, thereby allowing for flexible and dynamic reconstruction of expressions during computation.

\par This family of combinators, which we denote as \textbf{V}, specifies the positions within a tree where an element (the last argument) will be inserted. The multi-path is described using the following elements: \(<\) for \textit{Left}, \(>\) for \textit{Right}, and \(\langle , \rangle\) for \textit{Bifurcation}.

\begin{align*}
Path \,&= \, \, < \, Path \\
    &= \, \, > \, Path \\
    &= \, \langle Path_l,\, Path_r \rangle \\
    &= \varnothing
\end{align*}

\par We refer to the count of left \(<\) and right \(>\) elements as the \textit{degree} of a multi-path. Thus, the multi-path is effectively characterized by its degree, which represents the number of times each directional element appears within it.

\par To illustrate the concept of multi-paths and their degrees, consider the following examples:

\begin{itemize}
    \item \textbf{Example 1:} Multi-path \(< > <\) \\
    The multi-path \(< > <\) has a degree of 3, 2 for \((<\)) and 1 for (\(>\)).

    \item \textbf{Example 2:} Multi-path \(\langle \langle < , > \rangle ,  < \rangle\) \\
    This multi-path has a degree of 3, 2 for \(<\) and 1 for \(>\). The bifurcation \(\langle l, r \rangle\) indicates a branching, where the degree counts the total occurrences of \(<\) and \(>\) including those in the branches.
\end{itemize}

\par The intuition behind this family of combinators lies in its connection to the process of abstraction, particularly in the context of \((\lambda x . \, t)\), which abstracts the variable \(x\) from all its occurrences in a term \(t\). During the compilation process—specifically, in the abstraction elimination phase—each lambda abstraction is substituted by a corresponding combinator from this family. When this combinator is applied to a term from which all instances of \(x\) have been removed, along with a specific value, it effectively reinserts that value into all the positions where \(x\) was originally located prior to the abstraction.

\par The implementation of the \textbf{V} combinator presents a particular challenge, specifically in the layout of the tree argument. A naive approach would involve placing the tree as the first argument and the object to be inserted as the second argument of the \textbf{V} combinator. The expected result would be a tree where the object is inserted at all positions indicated by the multi-path directed by the \textbf{V} combinator.

\par However, such a combinator would not exhibit the optimal parallelism properties that standard combinators possess. In particular, it would require a more sequential evaluation, reducing the potential for simultaneous execution across multiple parts of the expression. This lack of inherent parallelism would limit the performance and scalability that our architecture is designed to achieve.

\par To solve this problem the tree gets laid out at the spine guided by the multipath.  The degree of the multi-path equals the number of arguments for the tree. Consequently, the arity of the \textbf{V} combinator is equal to the degree of the multi-path plus one (the additional argument for the element to be inserted).

\par For example, consider the lambda expression \(\lambda x. \, a \, (b \, (c \, x)) \, (x \, d)\). This expression would compile to the \textbf{V} combinator with a multi-path representation of \(\textbf{V}_{\langle > > > , < \rangle}\).

\begin{forest}
for tree={
    grow=south, 
    circle, draw, minimum size=1em, inner sep=2pt, s sep=9mm, 
    l=1cm 
}
[ 
    [ 
        [$a$] 
        [ 
            [$b$] 
            [ 
                [$c$] 
                [$x$] 
            ]
        ]
    ]
    [ 
        [$x$] 
        [$d$] 
    ]
]
\end{forest}

\par The compiler arranges the tree along the spine by following the specified path and treating the opposite side of the path as a spine element. For any bifurcation, the general approach is to traverse the left side first, followed by the right. This realized layout of arguments we call the residual.

\par In the given example, to write this tree as argument, we follow the path while we collect the arguments for the spine of the tree. Looking at the \textbf{V} combinator path we have an initial split, and we proceed down the left-hand branch, because we always do it first. Then we encounter \(a\) and right, then \(b\) and right, and finally \(c\) and right. So the current residual looks: \(a\) \(b\) \(c\).

But we have to do the right side as well, where \(d\) is encountered. So the spine residual is: \(a\) \(b\) \(c\) \(d\) and consequently, the compiler translates the expression \(\lambda x. \, a \, (b \, (c \, x)) \, (x \, d)\) into the combinator \(V_{\langle > > >, < \rangle} \, abcdx\), which evaluates to \(a \, (b \, (c \, x)) \, (x \, d)\).

\par We observe that the tree, into which the elements are to be inserted, is rearranged by the compiler such that the parts that will change at reduction are laid out along the spine. This provides the \textbf{V} combinator with optimal conditions for parallelism.

The \textbf{V} combinator, due to its complexity, may not exhibit a constant execution time. However, the architecture presented in the following section is designed to handle such variability efficiently. Unlike traditional reduction models that require one reduction per cycle, our machine is capable of accommodating the non-constant execution times of the \textbf{V} combinator without compromising performance.

\par Let us consider another example to further demonstrate the compilation process. Take the lambda expression:

\[
\lambda x. ( a \, b \, (x \, (c \, d))) \, (x \, (g \, h) \, (e \, f))
\]

\begin{forest}
for tree={
    grow=south, 
    circle, draw, minimum size=1em, inner sep=1.8pt, s sep=5mm, 
    l=1cm 
}
[ 
    [ 
        [ 
            [$a$] 
            [$b$] 
        ]
        [ 
            [$x$] 
            [ 
                [$c$] 
                [$d$] 
            ]
        ]
    ]
    [ 
        [ 
            [$x$] 
            [ 
                [$g$] 
                [$h$] 
            ]
        ]
        [ 
            [$e$] 
            [$f$] 
        ]
    ]
]
\end{forest}

\par To determine the multi-path for the expression \((a \, b \, (x \, (c \, d))) \, (x \, (g \, h) \, (e \, f))\), we analyze the positions where the variable \(x\) is applied. As observed, the multi-path for this expression is \(\langle ><, <<\rangle \).

\par In the given example, the initial steps follow the multi-path to arrange the residual. The first path, \(><\), proceeds by moving right and then left: first encountering \((a \,  b)\), followed by \((c \, d)\), and finally reaching \(x\). The second path, \(<<\), moves left twice encountering the subtrees \((e \, f)\) and \((g \, h)\).

\par Consequently, the compiler translates the expression \( \ x .(a \, b \, (x \, (c \, d))) \, (x \, (g \, h) \, (e \, f))\) into the combinator \(V_{\langle ><, << \rangle} \, (a \,  b) (c \, d) (e \, f) (g \, h) x\). This transformation ensures that the expression maintains its original structure, allowing the combinator to evaluate to the intended expression \((a \, b \, (x \, (c \, d))) \, (x \, (g \, h) \, (e \, f))\).

\subsection{The KVY Assembler Code}

\par The combinator base \textbf{KVY} is designed to provide a minimal yet powerful set of combinators that can represent any computation expressible in the lambda calculus. The base consists of three fundamental combinators: \textbf{K}, \textbf{V}, and \textbf{Y}.

\par As an example we take a program to calculate "one plus one". The numbers are represented as natural numbers. The program consists of the definition of the type of Nat, the definition of the function plus and the computation \textbf{plus 1 1}:

\begin{Verbatim}
\IdrisKeyword{data} \IdrisType{Nat} = \IdrisData{S} \IdrisType{Nat}  | \IdrisData{Z}
\IdrisFunction{plus} \IdrisData{Z} \IdrisBound{n} = \IdrisData{n}
\IdrisFunction{plus} (\IdrisData{S} \IdrisBound{m}) \IdrisData{n} = \IdrisData{S} (\IdrisFunction{plus} \IdrisBound{m} \IdrisBound{n})
\IdrisFunction{main} = \IdrisFunction{plus} 1 1
\end{Verbatim}

\par Compiled to KVY assembler, it looks this way:

\begin{Verbatim}
V<\{\{<>>>>>,\},\} (K V) V<< V<\{>>,>\}
Y (V<> V\{><>,\}) V<<> V><> (V>< K)
\end{Verbatim}

\par KVY has the following abstract reduction rules:

\begin{align*}
K \, x \, y \, &\Rightarrow \, x \\
V \, \varnothing \, w \, &\Rightarrow \, w \\
V \, ( < \, Path ) \, x \, \overline{x} \, w \, &\Rightarrow \, ( V \, Path \, \overline{x} \, w ) \, x \\
V \, ( > \, Path ) \, x \, \overline{x} \, w \, &\Rightarrow \, x \, ( V \, Path \, \overline{x} \, w ) \\
V \, \{ Path _l , \, Path _r \} \, \overline{x} _l \, \overline{x} _r \, w &\Rightarrow \, ( V \, Path _l \, \overline{x} _l \, w ) \, ( V \, Path _r \, \overline{x} _r \, w ) \\
Y \, f \, x \, &\Rightarrow \, f \, ( Y \, f ) \, x \\
\end{align*}

\par And this is the abstraction elimination algorithm for the KVY language:

\begin{align*}
\langle x \rangle _x \, &\mapsto \, \varnothing \\
\langle Tm _l \, Tm _r \rangle _x \, &\mapsto \, \, < \, \langle Tm _l \rangle _x \quad \text{if} \, x \, \notin \, Tm _r \\
\langle Tm _l \, Tm _r \rangle _x \, &\mapsto \, \, > \, \langle Tm _r \rangle _x \quad \text{if} \, x \, \notin \, Tm _l \\
\langle Tm _l \, Tm _r \rangle _x \, &\mapsto \, \, \{ \langle Tm _l \rangle _x \, , \, \langle Tm _r \rangle _x \} \\
\end{align*}

\begin{align*}
( x ) _x \, &\mapsto \, \cdot \\
( Tm _l \, Tm _r ) _x \, &\mapsto \, Tm _r \, ( Tm _l ) _x \quad \text{if} \, x \, \notin \, Tm _r \\
( Tm _l \, Tm _r ) _x \, &\mapsto \, Tm _l \, ( Tm _r ) _x \quad \text{if} \, x \, \notin \, Tm _l \\
( Tm _l \, Tm _r ) _x \, &\mapsto \, ( Tm _l ) _x \, ( Tm _r ) _x \\
\end{align*}

\begin{align*}
[\lambda \, x \, . \, Tm] \, &\mapsto \, K \, [Tm] \quad \text{if} \, x \, \notin \, Tm \\
[\lambda \, x \, . \, Tm] \, &\mapsto \, V \, \langle [Tm] \rangle _x \, ( [Tm] ) _x \\
[x] \, &\mapsto \, x \\
[Tm_l \, Tm_r] \, &\mapsto \, [Tm_l] \, [Tm_r] \\
[Y] \, &\mapsto \, Y
\end{align*}

\begingroup
\vspace*{-\baselineskip}
\captionof{figure}{KVY Abstraction Elimination}
\vspace*{\baselineskip}
\endgroup


\par The original work, on which the V combinator is based on, is from \cite{ClK}. Together, these combinators form a robust foundation for computation. The \textbf{KVY} base is minimal yet expressive, providing the necessary tools to represent complex computational constructs while ensuring that the resulting expressions are optimized for parallel evaluation.

\section{The Matrima machine}

\par The combinator code outlined in the last section necessitates a dedicated machine for its evaluation. In this section, we describe the architecture of such a machine. The ultimate objective of this project is to develop this machine as a specialized silicon chip, meticulously designed to execute combinator code with high performance. Although this goal has yet to be realized, we are currently conducting experiments using virtual machines that simulate the target architecture on contemporary CPUs. Additionally, we are exploring further steps to simulate the target architecture on GPUs and FPGAs, which may provide valuable insights into optimizing the machine's performance before its physical implementation.

\par The architecture must be meticulously designed to leverage the inherent advantages of functional programming in general and combinator code in particular. The success of this project hinges on the architecture’s ability to deliver high-speed performance for functional programs, coupled with greater energy efficiency compared to current CPUs. Furthermore, the architecture must facilitate the scalable deployment of these chips, ensuring that the system can be expanded easily to meet growing computational demands. Achieving these goals could position the Matrima Machine as a significant disruptive force within the current IT landscape, which is predominantly based on the Von Neumann architecture

\subsection{The Basics of the Matrima Machine}

\par At its core, the Matrima Machine represents a significant departure from traditional Von Neumann-based designs, which are characterized by their reliance on sequential instruction execution and shared memory models. In contrast, the Matrima Machine adopts a non-sequential, stateless computational model, which is inherently better aligned with the parallelism and immutability that are foundational to functional programming. This shift enables the architecture to fully exploit the advantages of functional code, particularly in terms of parallel processing efficiency and scalability.

\par The evaluation of combinator code necessitates two fundamental activities. The first activity involves determining whether an expression can be reduced, assessing whether the expression is in head normal form or in normal form. The second activity is the reduction process itself, wherein the expression is systematically simplified according to the rules of combinator logic.

\par One of the most significant advantages of combinator code lies in its inherent properties regarding parallelism. Specifically, any expression within combinator code can be reduced in parallel without any conflicts, which starkly contrasts with the sequential nature of the Von Neumann architecture. This characteristic of combinator code is the cornerstone of the Massimult architecture, and the design of the actual machine is meticulously crafted to exploit this advantage to its fullest extent. By maximizing parallel processing capabilities, the Matrima machine seeks to achieve superior performance and efficiency, setting it apart from traditional computational models.

\par The primary concept behind the Matrima machine is to assign a separate thread to every expression and subexpression, enabling all these threads to operate in parallel. Each thread is responsible for evaluating whether its respective expression can be reduced; if reduction is possible, the thread will carry out the reduction. To determine the reducibility of an expression, a thread needs only to inspect the relevant subexpressions.

\par As reductions occur, new expressions are generated, and corresponding threads are initiated to evaluate them. These threads also check whether the expression has reached head normal form, in which case it will not be further evaluated, or whether it has achieved normal form, in which case the expression, along with all its subexpressions, will no longer undergo any reductions.

\par Once an expression reaches normal form, no further threads are required for that expression. If the root cell of the computation reaches normal form, the overall computation is complete, and the result is ready for use. This approach ensures that the architecture efficiently leverages the parallelism inherent in combinator code, driving optimal performance.

\par While this concept of code evaluation is straightforward and simple in principle, it necessitates careful consideration of its implications. In this architecture, we are dealing with an extraordinarily large number of parallel threads. Each of these threads operates independently, without any communication or synchronization with other threads, yet collectively they contribute to producing a single correct result.

\par However, it is important to note that the order of these reductions can vary from one execution to another. As a result, a computation in this model does not consist of a predetermined sequence of operations; rather, it is the sum of all possible reduction orders, with a specific order being selected during each individual run. Consequently, when executing the identical program with identical inputs, two separate runs are highly likely to diverge in both the number of reductions performed and the amount of memory utilized. This variability introduces a degree of nondeterminism in resource usage, which must be carefully managed to ensure optimal performance and reliability.

\par A second, and perhaps even more unexpected, observation is that at any given moment during program execution, the global expression on which the threads operate may become inconsistent. This arises from the lack of synchronization, allowing reads and writes to occur in any arbitrary order. In our prototype implementation, we ultimately rely on operating system-level threads, which can be paused and resumed at times largely beyond our control, leading to certain writes being executed much later than anticipated. Given the absence of synchronization, we must acknowledge and accept this inherent inconsistency.

\par We anticipate that a similar situation will arise in the silicon-based implementation, as the architecture is deliberately designed to avoid synchronization or communication between threads, which would otherwise introduce latency and slow down the machine. This design choice, while potentially leading to temporary inconsistencies, is crucial for maintaining the high-speed performance and parallel efficiency that the Matrima Machine seeks to achieve.

\par This phenomenon of unpredictable timing manifests in two critical aspects of our prototype implementation. We employ simple reference counting for memory recycling. However, due to the asynchronous nature of the system, increments and decrements to reference counts can be written in any arbitrary order. This leads to the peculiar situation where a cell with a reference count of zero might later be referenced again, and in some cases, we even encounter negative reference counts.

\par We will revisit the issue of reference counting in more detail in the section dedicated to memory recycling. However, we can already anticipate that, as a consequence of this phenomenon, it is necessary to halt all reductions before initiating the recycling of cells. This precaution is essential to prevent the erroneous reclamation of memory that may still be in use, thereby ensuring the integrity of the computation.

\par The second unexpected phenomenon we encountered was related to normal forms. Initially, we assumed that when both the left and right subexpressions of any given expression are in normal form, and the expression itself cannot be further reduced, the expression must also be in normal form. However, due to the unpredictable timing of operations, we observed cases where reductions were written after the expression had already been tagged as being in normal form. This discrepancy led to inconsistencies in our evaluations.

\par To resolve this issue, we implemented an additional verification step, ensuring that the expression must also be in head normal form before it is conclusively tagged as being in normal form. This additional check mitigates the effects of timing unpredictability, thereby maintaining the accuracy and consistency of our normal form evaluations.

\par Despite the timing unpredictability inherent in our system, it remains true that the data is consistent at any moment when the evaluation is paused and all changes have been fully written. Additionally, it is important to emphasize that, regardless of the variations in timing, the correct result will always be computed. This consistency is a critical property of the Matrima Machine, ensuring that, even in the face of asynchronous operations and non-deterministic execution paths, the integrity of the computational process is preserved and reliable outcomes are guaranteed.

\par To summarize our approach: A vast number of threads operate in parallel, without any form of communication, synchronization, locking, or waiting on one another, yet they collaborate seamlessly to achieve a unified outcome. This is a rather remarkable achievement, demonstrating the robustness of the Matrima Machine's design. The ability to maintain coherence and correctness in the absence of traditional coordination mechanisms is both surprising and a testament to the power and efficiency of this parallel processing model.

\par However, there is one critical point of concurrency that remains: the allocation of fresh memory, which is a shared resource among all threads. To maintain the efficiency of the system, it is essential that memory allocation be implemented in such a way that it does not become a bottleneck. Specifically, we must avoid a scenario where there is a single point of allocation for all cells, which could result in threads being blocked while waiting for others to complete their allocations. Instead, memory allocation must be distributed or managed in a manner that allows threads to allocate memory independently, without being impeded by the activities of other threads.

\par Finally, in our prototype implementation, we have found it necessary to use atomic reads and writes of 128-bit chunks. This measure is essential to prevent inconsistencies that could arise, which are not merely due to timing unpredictabilities but could lead to critical errors in data integrity. By ensuring that these operations are atomic, we maintain consistency within the system, even in the absence of traditional synchronization mechanisms.

\par We will now provide a more detailed overview of the Matrima machine, a key component of the Massimult architecture, as illustrated in Figure 11. This section will delve into the specific components that underlies the Matrima machine, highlighting how it integrates within the broader architectural framework to achieve the desired performance and efficiency.

\includegraphics[width=0.5\textwidth]{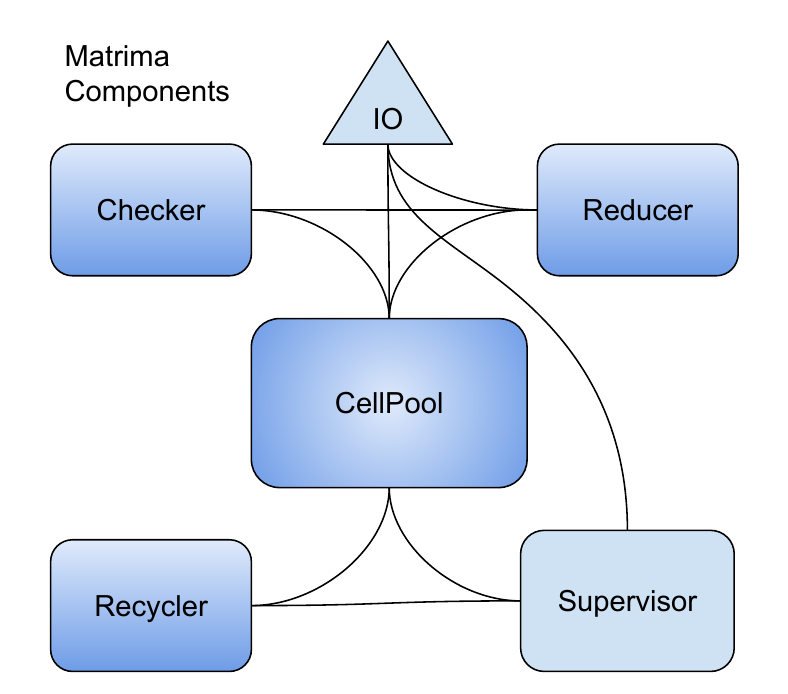}
\begingroup\vspace*{-\baselineskip}
\captionof{figure}[h]{Matrima Components}
\vspace*{\baselineskip}\endgroup

\section{Key Components of the Matrima Machine}

\subsection{The CellPool}
\par The \textbf{CellPool} represents expressions in memory.

\par The \textbf{CellPool} functions as an array of cells, specifically designed to facilitate the binary representation of combinator expressions within memory. In this context, a \textbf{cell} denotes the representation of either a node or a leaf within a combinator expression. The compiler is responsible for generating binary code, structured as an array of cells, which can be directly transferred into the CellPool without requiring any modifications.

\par For each Matrima process, the cells within the CellPool that are accessible from the process's root cell are maintained. Each cell in the CellPool is assigned a unique index, which serves as its reference point. As previously discussed, a computation is considered complete when the root cell reaches normal form. At this point, the result is produced as a binary array of cells, which is derived by compressing the cells, starting from the root cell, into a contiguous array of cells.

\par Each cell is currently represented as a 16-byte or 128-bit binary expression. This configuration provides an address space of \(2^{32}\) cells, allowing a CellPool to have a maximum size of \(2^{32}\) cells in this representation, which equates to approximately 69 gigabytes of memory. For applications requiring larger CellPools, it would be necessary to modify the current representation to accommodate the increased memory demands.

\par A cell consists of 64 bytes of content, a 16-byte reference count, 16-bit flags, 16-bit checker data, and 16 bits reserved for future use. A cell can either be \textit{alive} or \textit{garbage}. If a cell is classified as garbage, it does not represent any meaningful data but remains available for future allocations. Conversely, if a cell is alive, it can be a node or a leaf.

\par In the case where a cell is a node, its content part consists of 32-bit left and right components. These typically serve as indices within the CellPool, representing combinator graphs. As an optimization, the content part can also contain embedded combinators or primitives, a configuration specified by the cell's flags. Additional flags for nodes store information regarding whether a node is in head normal form or normal form. These flags are crucial for the evaluation process, as they help determine the state of a node's reducibility and guide the reduction strategy accordingly.

\par If a cell is classified as a leaf, it may represent either combinators or primitives, where primitives can include both primitive operations and primitive data. The flags associated with leaves provide a description of all primitive data types, ensuring that each leaf is properly identified and processed according to its specific type.

\par It is important to note that the reference count may become negative due to the unpredictable timing issues previously described. To accommodate this, the reference count is represented as \texttt{refcount + 16}, allowing for the necessary range to account for potential negative values. The 16 bits allocated for checker arity will be discussed in detail in the subsequent section.

\begin{table}[!ht]
        \centering

\begin{tabular}{p{0.03\textwidth}p{0.15\textwidth}p{0.15\textwidth}}
Byte & Node & Leaf \\
\hline
 \multicolumn{1}{|l|}{0} & \multicolumn{1}{l|}{} & \multicolumn{1}{l|}{} \\
\cline{1-1}
 \multicolumn{1}{|l|}{1} & \multicolumn{1}{l|}{Left child} & \multicolumn{1}{l|}{} \\
\cline{1-1}
 \multicolumn{1}{|l|}{2} & \multicolumn{1}{l|}{} & \multicolumn{1}{l|}{} \\
\cline{1-1}
 \multicolumn{1}{|l|}{3} & \multicolumn{1}{l|}{} & \multicolumn{1}{l|}{Contents} \\
\cline{1-2}
 \multicolumn{1}{|l|}{4} & \multicolumn{1}{l|}{} & \multicolumn{1}{l|}{} \\
\cline{1-1}
 \multicolumn{1}{|l|}{5} & \multicolumn{1}{l|}{Right child} & \multicolumn{1}{l|}{} \\
\cline{1-1}
 \multicolumn{1}{|l|}{6} & \multicolumn{1}{l|}{} & \multicolumn{1}{l|}{} \\
\cline{1-1}
 \multicolumn{1}{|l|}{7} & \multicolumn{1}{l|}{} & \multicolumn{1}{l|}{} \\
\hline
 \multicolumn{1}{|l|}{8} & \multicolumn{1}{l|}{Flags} & \multicolumn{1}{l|}{Flags} \\
\cline{1-1}
 \multicolumn{1}{|l|}{9} & \multicolumn{1}{l|}{} & \multicolumn{1}{l|}{} \\
\hline
 \multicolumn{1}{|l|}{10} & \multicolumn{1}{l|}{Reference count + 16} & \multicolumn{1}{l|}{Reference count + 16} \\
\cline{1-1}
 \multicolumn{1}{|l|}{11} & \multicolumn{1}{l|}{} & \multicolumn{1}{l|}{} \\
\hline
 \multicolumn{1}{|l|}{13} & \multicolumn{1}{l|}{Checker arity} & \multicolumn{1}{l|}{Reserved} \\
\cline{1-1}
 \multicolumn{1}{|l|}{12} & \multicolumn{1}{l|}{} & \multicolumn{1}{l|}{} \\
\hline
 \multicolumn{1}{|l|}{13} & \multicolumn{1}{l|}{Reserved} & \multicolumn{1}{l|}{Reserved} \\
\cline{1-1}
 \multicolumn{1}{|l|}{15} & \multicolumn{1}{l|}{} & \multicolumn{1}{l|}{} \\
 \hline
\end{tabular}
\end{table}

\par The CellPool can be conceptualized as the memory of the processor. However, as will be discussed in the next section, it functions as active memory, in that every cell can concurrently execute the checker task. This parallel capability allows the architecture to efficiently manage computations and optimize performance.

\subsection{The Checker}
The \textbf{Checker} identifies reducible expressions, head normal forms, and normal forms.

\par The checker is responsible for evaluating three key aspects of a node: (1) determining whether the node can be reduced, (2) assessing if the node is in head normal form, and (3) verifying if the node is in normal form. The checker should be implemented in hardware to enable simultaneous operation on all active node cells in parallel.

\par To determine if a cell can be reduced, the checker traverses down the left path of the expression until a combinator or primitive operator is encountered. A cell is deemed reducible if the arity of the identified combinator or primitive operator matches the number of steps taken during this traversal.

\par However, we employ an alternative implementation that utilizes the 16-bit checker arity of each node cell to determine its reducibility. This approach works by propagating the arity upwards step by step while decrementing it by 1 at each step. In practice, each cell reads the arity value of its left child. If the left child is a leaf containing a combinator or primitive operation, the checker arity is set to the arity of the combinator or primitive minus 1. If the left child is another node, the checker arity is set to the checker arity of that node minus 1. A cell is considered reducible when its checker arity reaches 0.

\par To determine if a cell is in head normal form, it is sufficient to check whether the checker arity is greater than zero. In this case this cell will never reduce.

\par The third task of the checker is to determine whether a node cell is in normal form. A node cell is considered to be in normal form if it is in head normal form and both its left and right children are also in normal form. When a cell is in normal form, neither the cell itself nor any of its subterms can be further reduced in the future. This concludes the description of the checker's functionality.

\subsection{The Reducer}
The \textbf{Reducer} performs reductions on single combinators or primitives.

\par The reducer performs a destructive update on the top cell of a reducible expression to ensure that the cell correctly represents the result of the reduction. During this process, the reducer may allocate new cells, which subsequently initiate new checker tasks. Additionally, the reducer adjusts the reference counts, which may lead to certain cells becoming unreferenced and, therefore, no longer participating in representing any expression. Finally, the reducer sets the checker arities for the newly allocated cells to facilitate subsequent evaluations.

\par The reducer may also perform primitive operations, such as the division of floating-point numbers or input-output operations. Both types of operations involve interaction with the Supervisor component: the former may trigger a "division by zero" error, while the latter may cause the process to block while waiting for input.

\par The reducer serves as the core component of the Matrima machine. The description provided here is concise, as the details of combinators have already been covered in Part 1, and the specifics of primitive operations fall outside the scope of this paper.

\subsection{The Recycler}
The \textbf{Recycler} handles memory recycling for reuse, ensuring efficient memory management.

\par As previously discussed, the order of updates in the system is non-deterministic. This characteristic has implications for reference counting: it can result in negative reference counts, or reference counts that have reached zero may become positive again. To manage this, we halt the evaluation process prior to recycling and resume it afterward. Recycling involves returning a cell with a reference count of zero to the pool of allocable memory. If the cell is a node, the reference counts of its left and right subcells are decremented accordingly. This is a recursive process, as the reference counts of these subcells may also reach zero, necessitating further recycling. Once this process is complete, the reduction can be resumed.

\subsection{The Supervisor}
The \textbf{Supervisor} manages computational processes, overseeing the coordination and execution of tasks within the architecture.

\par A single CellPool can be utilized to perform multiple computations concurrently by simply designating different root cells for each computation. There is no sharing of cells between different processes. Each process is managed by a dedicated supervisor, which is responsible for loading the code into the CellPool, maintaining a record of the root cell, detecting when the computation has concluded, and subsequently compressing and returning the result.

\par The supervisor is also responsible for handling runtime errors, such as division by zero or out-of-memory conditions. Additionally, it collaborates with input-output primitives to monitor the state of the process, which may, for example, be waiting for a specific input. The supervisor's introspective capabilities will further encompass data metrics such as the time and memory utilized by the process.
\section{Future Work and Conclusion}

\par The Massimult architecture presents a promising new direction for building highly parallel, energy-efficient, and scalable computing systems based on combinator reduction. While this work has demonstrated the theoretical foundations and implemented a virtual machine simulation, several key areas require further exploration and development to fully unlock the potential of this approach.

\subsection{Future Work}
Several key areas for future research and development are outlined below:

\begin{itemize}
    \item \textbf{GPU and FPGA Implementation}: To gain traction and showcase the potential of this architecture, it is crucial to demonstrate that it can execute certain programs quicker and with reduced memory consumption on current hardware. The immediate goal is to approach a GPU implementation, followed by targeting FPGAs. These platforms offer the opportunity to scale the architecture and evaluate its performance in real-world high-performance computing environments.

    \item \textbf{Optimization in Speculative Evaluation (Compile Time)}: One of the critical areas for improvement lies in optimizing the speculative evaluation strategy at compile time. By enhancing the compiler's ability to predict which parts of a program will likely need evaluation, unnecessary computations can be minimized, further enhancing the system's performance.

    \item \textbf{Seamless Integration of Array Structures (Compile Time and Runtime)}: Initial research on integrating array structures into the combinator model has yielded promising results. Future work will focus on seamlessly supporting arrays at both compile time and runtime, ensuring efficient handling of indexed data while preserving the benefits of combinator reduction.

    \item \textbf{Compiling Functional Languages to LambdaB and LambdaM}: To execute a wider range of real-world functional programs using the Massimult architecture, one of our intermediate languages shall serve as a backend language for existing functional languages such as Idris. Compiling these languages to an intermediate representation will expand the usability and adoption of the Massimult architecture, fostering a broader and more diverse ecosystem.

    \item \textbf{Localization in Cell Pools (Runtime)}: Enhancing memory allocation and management is another essential aspect of future work. By localizing the allocation of cells in cell pools, locality can be optimized, leading to improved performance during runtime. Additionally, for long-running processes, an automatic reordering of cells should be implemented to further enhance performance over time.

    \item \textbf{Hardware Implementation and Development}: The ultimate objective is the design and fabrication of a physical chip based on the Massimult architecture. This hardware will fully exploit the inherent parallelism and energy efficiency of combinator reduction. After initial validation, the project will progress toward developing a full-scale hardware prototype, which will be essential for benchmarking performance and energy efficiency gains. This step holds the potential to fundamentally transform modern CPU design.
\end{itemize}

\subsection{Conclusion}
\par The Massimult architecture represents a fundamental departure from traditional Von Neumann-based computing systems. By leveraging the parallelism inherent in lambda calculus and combinator reduction, Massimult offers the potential for more efficient computation, reduced energy consumption, and improved scalability. Although our current implementation remains in its early stages, it has demonstrated the feasibility of this approach and laid a strong foundation for future advancements in both software and hardware.

\par Ultimately, the success of the Massimult architecture will depend on demonstrating its advantages in real-world applications, particularly in terms of performance and energy efficiency. If these benefits are realized, the Massimult architecture could disrupt the current landscape of computing, providing a foundation for future systems that are faster, more efficient, and more scalable than ever before.

\section{Appendix}



\nocite{*}
\printbibliography

\end{document}